# Compton scattering for photon and gluon in fixed-target collisions at AFTER@LHC

Gongming Yu,[1] Runlong Liu,[1] Yanbing Cai,[2, *] Quangui Gao,[3] and Qiang Hu[4]

[1] Fundamental Science on Nuclear Safety and Simulation Technology Laboratory,
Harbin Engineering University, Harbin 150000, China
[2] Guizhou Key Laboratory in Physics and Related Areas,
And Guizhou Key Laboratory of Big Data Statistic Analysis,
Guizhou University of Finance and Economics, Guiyang 550025, China
[3] Department of Physics, Yuxi Normal University, Yuxi 653100, China
[4] Institute of Modern Physics, Chinese Academy of Sciences, Lanzhou 730000, China

We calculate the Compton scattering for photon and gluon with the Klein-Nishina formula in fixed-target collisions by using the proton and lead beams at AFTER@LHC. In these collisions, we can investigate the particular case of Compton scattering at the partonic level, such as γq→qγ, gq→qγ, γq→qg, and gq→qg, that can help to check of the equivalent-photon approximation and understand the dynamics of hadron collisions at high energies, as well as probe the inner hadron structure.

## I. INTRODUCTION

Fixed-target collisions permit the study of photon, lepton, jet, and hadron production in the target fragmentation region, as well as the structure of nuclear matter and the spin composition of a nucleon. Indeed, many phenomenological models predict that a first-order phase transition may occur in compressed baryon-rich matter created in the fixed-target collisions with low energy [1-9]. In this context, we find it important to study the potentialities offered by fixed-target program in High Intensity Heavy Ion Accelerator Facility (HIAF) [10], Compressed Baryonic Matter (CBM) experiment at Facility for Antiproton and Ion Research (FAIR) [11,12], Multi-Purpose Detector (MPD) experiment at Nuclotron-based Ion Collider fAcility (NICA) [13], Beam Energy Scan (BES) program at Relativistic Heavy Ion Collider (RHIC) [14-16], and AFTER@LHC at large hadron collider (LHC) [17]. Especially for the AFTER@LHC at LHC, the System for Measuring Overlap with Gas (SMOG) [18] would become a quarkonium, prompt photon, dilepton, jet, and heavy-flavour observatory since its large expected luminosity for a relatively high center-of-mass system (c.m.s.) energy of 115 GeV per nucleon with a 7 TeV proton beam and 72 GeV per nucleon with a 2.76 ATeV lead beam, with the high precision typical of the fixed-target mode [19-24]. In such collisions, one can investigate specific reactions, such as Compton scattering, where one of the colliding particle serves as a emitter of a photon or gluon and the other serves as a target.

The scattering between photons and electrons is one of the most important physical processes that can generically be called Compton scattering noticed by A. H. Compton in 1923 [25,26]. The wavelength shift of the scattered photon, $\Delta\lambda = (\hbar/mc)(1-\cos\theta)$ was firstly observed in Compton scattering. In this process, all servable phenomena involve photon-electron interactions and convincingly demonstrated that light comprises particles with energy and momentum. The calculation of the Compton scattering process by Dirac [27] and Gordon [28], and with full spin and relativistic corrections by Klein and Nishina [29], provided a convincing case of the Dirac equation. Indeed, the scattering of photon or gluon on matter through Compton scattering is a powerful tool to study its inner structure. Because of the widely applications, many efforts have been made to develop theoretical methods on ab initio calculations for Compton scattering process. Several approaches, such as the free electron approximation (FEA) [29,30], impulse approximation (IA) [31-34], incoherent scattering function/incoherent scattering factor (ISF) [35-37], and scattering matrix (SM) [38-46] are developed. In the present work, we study the Compton scattering for photon and gluon with the Klein-Nishina formula [29] in fixed-target collisions by using the proton and lead beams at AFTER@LHC. In such collisions, we can investigate the particular case of Compton scattering at the partonic level, such as γq→qγ and γq→qg, for the proton and lead beam running on the proton-target. In the proton-target rest frame, the energy of photons can become significant if the energy of the moving charge (the proton and lead beam energy), becomes ultra-relativistic, as at the LHC. In photon-hadron collisions, relativistically moving charged hadrons are accompanied by electromagnetic fields that can effectively be used as quasi-real-photon beams obtained from a semi-classical description of high-energy electromagnetic collisions. At very high energies, these quasi-real-photons are energetic enough to initiate hard interactions Indeed, the production of gluon from Compton scattering is also very interesting process because it can help to understand the dynamics of hadron collisions at high energies. It can test the calculation of perturbative quantum chromodynamics (pQCD), and probe the quark matter.

In this paper, we report on a feasibility study of Compton scattering for photon and gluon at fixed-target collisions at AFTER@LHC using LHC beams. In Sec. II we present the

---

* myparticle@163.com

Compton scattering for photon and gluon in fixed-target collisions for the proton and lead beam running on the proton-target at AFTER@LHC. The numerical results for photon and gluon from Compton scattering in p-p collisions and p-Pb collisions at AFTER@LHC energies are plotted in Sec. III. Finally, the conclusion is given in Sec. IV.

## II. GENERAL FORMALISM

We study photon and gluon emission by relativistic heavy ions. When traversing a proton target, the projectile proton or ions could interact with the proton-target. Based on Klein-Nishina formula [29], the quark in Compton scattering are treated as free quark in the laboratory system (proton-target), all binding effects and many-body interactions are neglected in the scattering process. The prompt photon and gluon can be produced by the initial photon and gluon interacting with the quark from proton-target, such as $\gamma q \to q\gamma$, $gq \to q\gamma$, $\gamma q \to qg$, and $gq \to qg$ processes. In this situation, the factorized cross section of the Compton scattering for photon and gluon in fixed-target collisions for the proton and lead beam running on the proton-target at AFTER@LHC, can be written as

$$\sigma(pA \to \gamma X) = \int dx f_{q/p}^{Target} f_{\gamma/N}(x) \hat{\sigma}(\gamma q \to q\gamma) \\ + \int dx f_{q/p}^{Target} f_{g/N}(x) \hat{\sigma}(gq \to q\gamma),$$ (1.1)

$$\sigma(pA \to gX) = \int dx f_{q/p}^{Target} f_{\gamma/N}(x) \hat{\sigma}(\gamma q \to qg) \\ + \int dx f_{q/p}^{Target} f_{g/N}(x) \hat{\sigma}(gq \to qg),$$ (1.2)

where $f_{q/p}^{Target}$ is the quark distribution of the proton-target. In the laboratory system at rest, we chose the the quark distribution as (uud) for the proton-target in the quark model [47,48]. $x$ is the momentum fraction of photon (or gluon).

Based in the Klein-Nishina formula, the total cross section for Compton scattering $\hat{\sigma}(\gamma q \to q\gamma)$, $\hat{\sigma}(gq \to q\gamma)$, $\hat{\sigma}(\gamma q \to qg)$, and $\hat{\sigma}(gq \to qg)$ can be written as

$$\hat{\sigma}(\gamma q \to q\gamma) = \frac{4\pi\alpha^2 e_q^4}{6m_q^2}\left(\frac{\omega_f}{\omega_i}\right)^2\left(\frac{\omega_f}{\omega_i} + \frac{\omega_i}{\omega_f} - \frac{2}{3}\right),$$ (1.3)

$$\hat{\sigma}(gq \to q\gamma) = \frac{4\pi\alpha\alpha_s e_q^2}{16m_q^2}\left(\frac{\omega_f}{\omega_i}\right)^2\left(\frac{\omega_f}{\omega_i} + \frac{\omega_i}{\omega_f} - \frac{2}{3}\right),$$ (1.4)

$$\hat{\sigma}(\gamma q \to qg) = \frac{4\pi\alpha\alpha_s e_q^2}{3m_q^2}\left(\frac{\omega_f}{\omega_i}\right)^2\left(\frac{\omega_f}{\omega_i} + \frac{\omega_i}{\omega_f} - \frac{2}{3}\right),$$ (1.5)

$$\hat{\sigma}(gq \to qs) = \frac{4\pi\alpha_s^2}{6m_q^2}\left(\frac{\omega_f}{\omega_i}\right)^2\left(\frac{\omega_f}{\omega_i} + \frac{\omega_i}{\omega_f} - \frac{2}{3}\right),$$ (1.6)

where $\omega_f$ is the energy of final photon or gluon, $\omega_i = xE_{beam}$ is the energy of the initial photon or gluon, the beam energy $E_{beam}$ is 7TeV and 13TeV for proton beam, as well as 2.76TeV and 5.02TeV for lead beam at LHC. In the laboratory system at rest, the momentum of the quark in the proton-target is $p=(m_q,0,0,0)$, where $m_q$ is the mass of valence quark in proton.

The equivalent photon spectrum for charged nucleus moving with a relativistic factor $\gamma \gg 1$, can be obtained from the semiclassical description of high-energy electro-magnetic collisions. It will attain very large values since the nuclear charge then acts as a whole. The form factor is still not zero at $Q^2 \sim 1/b^2 \sim 1/R^2$, the photon spectrum of a pointlike nucleus with a cutoff at $Q^2 \sim 1/R^2$ was used which can equivalent to neglecting the nuclear size for high-energy nuclear collisions, where $R$ is the rdius of the nucleus and b is the impact parameters. A relativistic nucleus with the electric charge Ze moving with a relativistic factor $\gamma \gg 1$ with respect to some observer develops an equally strong magnetic-field component. It resembles a beam of real photons where the photon spectrum function of low photon energies can be written as [49-52]

$$f_{\gamma/N}(\omega) = \frac{2Z^2\alpha}{\pi\omega}\ln\frac{\gamma}{\omega R},$$ (1.7)

where $\omega$ is the momentum of photon and $R=b_{min}$ is the radius of the nucleus ($b_{min}$ is the cutoff of impact parameters). In the logarithmic approximation, the results can obtained from a purely classical treatment or by including the form factor are related to each other through a rescaling of the relativistic factor.

For proton, the equivalent photon spectrum function can be obtained from the Weizsacker-Williams approximation [53-55],

$$f_{\gamma/p}(x) = \frac{\alpha}{2\pi x}\left[1+(1+x)^2\right] \\ \times \left(\ln A_p - \frac{11}{6} + \frac{3}{A_p} - \frac{3}{2A_p^2} + \frac{1}{2A_p^3}\right),$$ (1.8)

where α is the momentum fraction of photon, $A_p=1+0.71 \text{GeV}^2/Q_{min}^2$ with

$$Q_{min}^2 = -2m_p^2 + \frac{1}{2s}\Big\{(s+m_p^2)(s-xs+m_p^2) \\ -(s-m_p^2)\sqrt{(s-xs+m_p^2)^2 - 4m_p^2 xs}\Big\},$$ (1.9)

here $m_p$ is the mass of proton, and at high energies $Q_{min}^2$ is given to a very good approximation by $m_p^2 x^2/(1-x)$.

The gluon distribution for nucleus $f_{g/N}(x)$ is given by [56,57]

$$f_{g/N}(x) = R_A(x)A[xg(x)],$$ (1.10)

where $R_A(x)$ is the nuclear modification factor [58], A is the nucleon number of the nucleus. The factor $xg(x)$ is the gluon

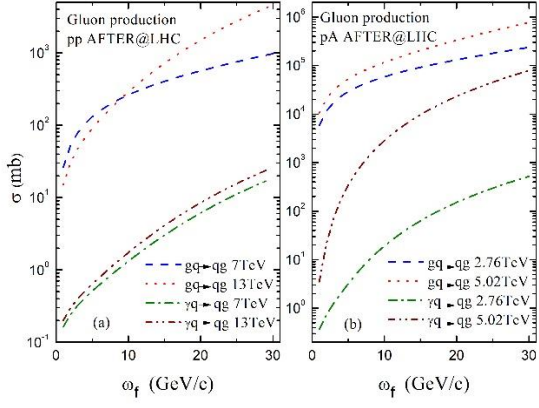

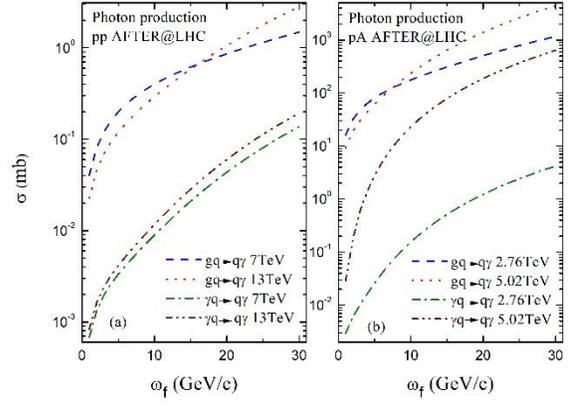

FIG. 1. The total cross section of gluon production from Compton scattering for the proton and lead beam running on the proton-target at AFTER@LHC. The dashed line (blue line) is for gq→qg by the proton (lead) beam with $E_{beam}=7\,\text{TeV}$ ($E_{beam}=2.76\,\text{ATeV}$) running on the proton-target, the dotted line (red line) is for gq→qg by the proton (lead) beam with $E_{beam}=13\,\text{TeV}$ ($E_{beam}=5.02\,\text{ATeV}$) running on the proton-target, the dashed-dotted line (green line) for γq→qg by the proton (lead) beam with $E_{beam}=7\,\text{TeV}$ ($E_{beam}=2.76\,\text{ATeV}$) running on the proton-target, the dashed-dotted-dotted line (dark red line) is for or γq→qg by the proton (lead) beam with $E_{beam}=13\,\text{TeV}$ ($E_{beam}=5.02\,\text{ATeV}$) running on the proton-target.

FIG. 2. The same as Fig. 1 but for photon production from Compton scattering for the proton and lead beam running on the proton-target at AFTER@LHC.

distribution function of nucleon, that can be parametrized by the functional form [59,60]

$$xg(x) = A_0 x^{A_1}(1-x)^{A_2}, \quad (1.11)$$

here, the free parameters $A_0=30.4571$, $A_1=0.5100$, and $A_2=2.3823$ are fixed by a global analysis of both the total cross-section data below medium energy [60].

## III. NUMERICAL RESULTS

We present the calculations of Compton scattering spectra for photon and gluon for proton and lead beam incident on a fixed proton-target. Figs. 1 and 2 show the spectrum of gluon and photon production for proton and lead ions aimed at a proton target, respectively. The contribution for the quark-gluon Compton scattering is important since the high density gluons of the nucleons. Seen from proton-target at rest, the photon spectrum becomes important. Especially for the nucleus, the equivalent photon spectrum obtained from semiclassical description of high-energy electromagnetic collisions is $f_{\gamma/N} \sim Z^2 \ln\gamma$, cross sections are enhanced by a factor of $Z^2$ and the relativistic factor becomes very LHC energies.

Therefore the contribution of photon and gluon production from photonquark Compton scattering is evident at AFTER@LHC. But for the proton beam running on the proton-target, the contribution for the photon-quark Compton scattering is small comparing with gluon-quark Compton scattering process.

## IV. CONCLUSIONS

In summary, we have investigated the production of photon and gluon from the Compton scattering with the Klein-Nishina formula for the proton and lead beam-s running on the fixed proton-target at AFTER@LHC. All binding effects and many-body interactions are neglected in the scattering process in the laboratory system (proton-target), and the quarks of proton-target are treated as free quark. In such collisions, we have investigated the particular case of gluon-quark Compton scattering at the partonic level, such as gq→qγ and gq→qg, as well as photon-quark Compton scattering (γq→qγ and γq→qg) since the equivalent photon spectrum from semiclassical description of high-energy electromagnetic collisions can become significant at LHC energies. Therefore, the Compton scattering for photon and gluon are important for us to check of the equivalent-photon approximation and understand the dynamics of hadron collisions at high energies, as well as probe the inner hadron structure.

## ACKNOWLEDGMENTS

This work is supported by Heilongjiang Science Foundation Project under Grant No. LH2021A009, and National Natural Science Foundation of China (No. 12063006).